\documentclass{aa}  

\usepackage{graphicx}
\usepackage{txfonts}
\def\msun{M$_\odot$}

\begin{document}

   \title{Orbit determination for the binary Cepheid V1344~Aql}

   \author{B. Cseh\inst{1,2,3}\fnmsep\thanks{cseh.borbala@csfk.hun-ren.hu},
          G. Cs\"ornyei\inst{1,2,4,5},
          L. Szabados\inst{1,2},
          B. Cs\'ak\inst{1,2},
          J. Kov\'acs\inst{3,6,7},
          L. Kriskovics\inst{1,2},
          A. P\'al\inst{1,2}
          }

   \institute{HUN-REN Research Centre for Astronomy and Earth Sciences, Konkoly Observatory, Konkoly Thege Mikl\'os út 15-17., H-1121, Hungary
         \and
    CSFK, MTA Centre of Excellence, Budapest, Konkoly Thege Mikl\'os út 15-17., H-1121, Hungary
         \and
    MTA--ELTE Lend{\"u}let ``Momentum'' Milky Way Research Group, Hungary
        \and
    Max-Planck-Institute for Astrophysics, Karl-Schwarzschild-Str. 1, 85741 Garching, Germany
		\and
	Technical University Munich, TUM Department of Physics, James-Franck-Str. 1., 85741 Garching, Germany 
        \and
    ELTE E\"otv\"os Lor\'and University, Gothard Astrophysical Observatory, 9700 Szombathely, Szent Imre H. str. 112, Hungary
        \and
    MTA-ELTE Exoplanet Research Group, 9700 Szombathely, Szent Imre H. str. 112, Hungary
    }
             
   \titlerunning{Orbit determination for the Cepheid V1344~Aql}
   \authorrunning{Cseh, Cs\"ornyei, Szabados et~al.}
   \date{Received; accepted}

 
  \abstract
   {Binary Cepheids play an important role in investigating the calibration of the classical Cepheid period-luminosity relationship. Therefore a thorough study of individual Cepheids belonging to binary systems is necessary.}
   {Our aim is to determine the orbit of the binary system V1344~Aql using newly observed and earlier published spectroscopic and photometric data.}
   {We collected new radial velocity observations using medium resolution (${R \approx 11000}$ and ${R \lessapprox 20000}$) spectrographs and we updated the pulsation period of the Cepheid based on available photometric observations using $O-C$ diagram. Separating the pulsational and orbital radial velocity variations for each observational season (year), we determined the orbital solution for the system using $\chi^2$ minimisation.}
   {The updated pulsation period of the Cepheid estimated for the epoch of HJD 2458955.83 is 7.476826 days. We determined orbital elements for the first time in the literature. The orbital period of the system is about 34.6 years, with an eccentricity $e$ = 0.22.}
   {}


   \maketitle
\section{Introduction}

The study of classical Cepheids (hereafter Cepheids) in binary (or multiple) systems plays an important role in the calibration of their well-known period-luminosity relationship \citep{Evans92, Breuval20, Karczmarek23}.
Unrevealed companion stars contribute to both the colour and the brightness of the Cepheid \citep{Szabados13}, falsifying the relationship \citep{SzabadosKlagyivik12, Gaia17}.
Additionally, the constantly growing number of binary Cepheids with known orbital elements can improve constraints on stellar evolution modelling \citep{Neilson15, Moe17, Karczmarek22}. 

The minimum orbital period of a binary system with a supergiant Cepheid is approximately one year \citep{Evans13, Szabados13, Neilson15}.
Shorter periods could be observed for Cepheids crossing the instability strip for the first time.
Such a system is found by \citet{Pilecki22} in the Large Magellanic Cloud, with an orbital period of 59 days. 
Short orbital period binary systems with a Cepheid component are not only important because of evolutionary studies, but also to test possible binary interactions and the birth of Cepheids through non-canonical evolution. 
Long-term radial velocity (RV) monitoring is necessary to identify Cepheids in a binary system with decades-long orbits and to have a more comprehensive picture of the fraction of binary Cepheids in different types of galaxies at a given metallicity. 

\citet{Szabados12_LMC} studied binarity of Cepheids in the Magellanic Clouds and presented a list of known binaries in both the Large and Small Magellanic Clouds with 17 and 15 pairs of stars, respectively. 
They concluded that the lack of known binaries is due to missing observational data and suggested more follow-up RV measurements.
\citet{Kervella19} searched for binaries among Galactic Cepheids and RR Lyrae type variables showing proper motion anomaly using {\it{Hipparcos}} and {\it{Gaia}} DR2 data. 
They found that the binary fraction of classical Cepheids might exceed 80\%, while \citet{Evans15} estimated it to be around 29\% based on stars with an orbital period below 40 years (and even lower for longer period systems, see their Table~7). 
This outstanding growth in the estimated binary fraction of binary Cepheids urges to more RV observations in order to determine orbital properties. 
The online database of Galactic binary Cepheids\footnote{available at \url{https://konkoly.hu/CEP/intro.html}} \citep{Szabados03} currently contains 171 Cepheids (including five systems awaiting confirmation) and among these 30 Cepheids with known orbital elements.


\citet{Szabados14} discovered the binary nature of V1344~Aql and suggested a relatively short, decades long orbital period. 
However, the available RV data were
insufficient to determine the period or other orbital parameters for the system.
Here we update the pulsation period and present new RV data for the binary Cepheid V1344~Aql and determine an orbital solution for the system.
In Section \ref{sec:period} we describe the updated pulsation period for the Cepheid. 
In Section \ref{sec:rvdata} we show our new RV data and the method of the orbit determination, respectively. 
We discuss our results in Section \ref{sec:summary}.
   
\section{Updated pulsation period}
\label{sec:period}
To derive the pulsation period valid at the time of observations, we updated the $O-C$ diagram presented in \cite{Csornyei2022}. This $O-C$ diagram showed a slow period increase in the time interval of 1975--2019. However, due to the large scatter among the $O-C$ values and the relatively short temporal coverage of the diagram, the period change rate could only be established with large uncertainties (${\dot{P} = 2.002\cdot 10^{-4} \pm 2.334\cdot 10^{-4}\,{\rm d/(100~yr)}}$, \citealt{Csornyei2022}). To update this value, we included more recent $V$ band observations from the Kamogata-Kiso-Kyoto Wide-field Survey (KWS, \citealt{KWS}). This includes about 200 additional data points in the timeframe of 2019--2023 to the previous data set, extending the coverage of the $O-C$ diagram by four years. Similarly, as in \cite{Csornyei2022}, the $O-C$ values were estimated for the median brightness value on the rising branch, following the reasoning described in \cite{Derekas2012}. For the individual $O-C$ values the data sets were split into 350 days long bins and these were fitted separately, while their uncertainties were estimated via bootstrapping. For the calculation of the epoch of median brightness ($T_{\textrm{med}}$, measured in HJD), the following elements were used:

\begin{equation}
\begin{split}
\label{eq:oc}
T_{\textrm{med}} = 2458955.8254 + 7.4768263 \textrm{ days}\\
 \pm 0.000006 \textrm{ day}
\end{split}
\end{equation}

\begin{figure*}[!ht]
\centering
\includegraphics[width=0.7\linewidth]{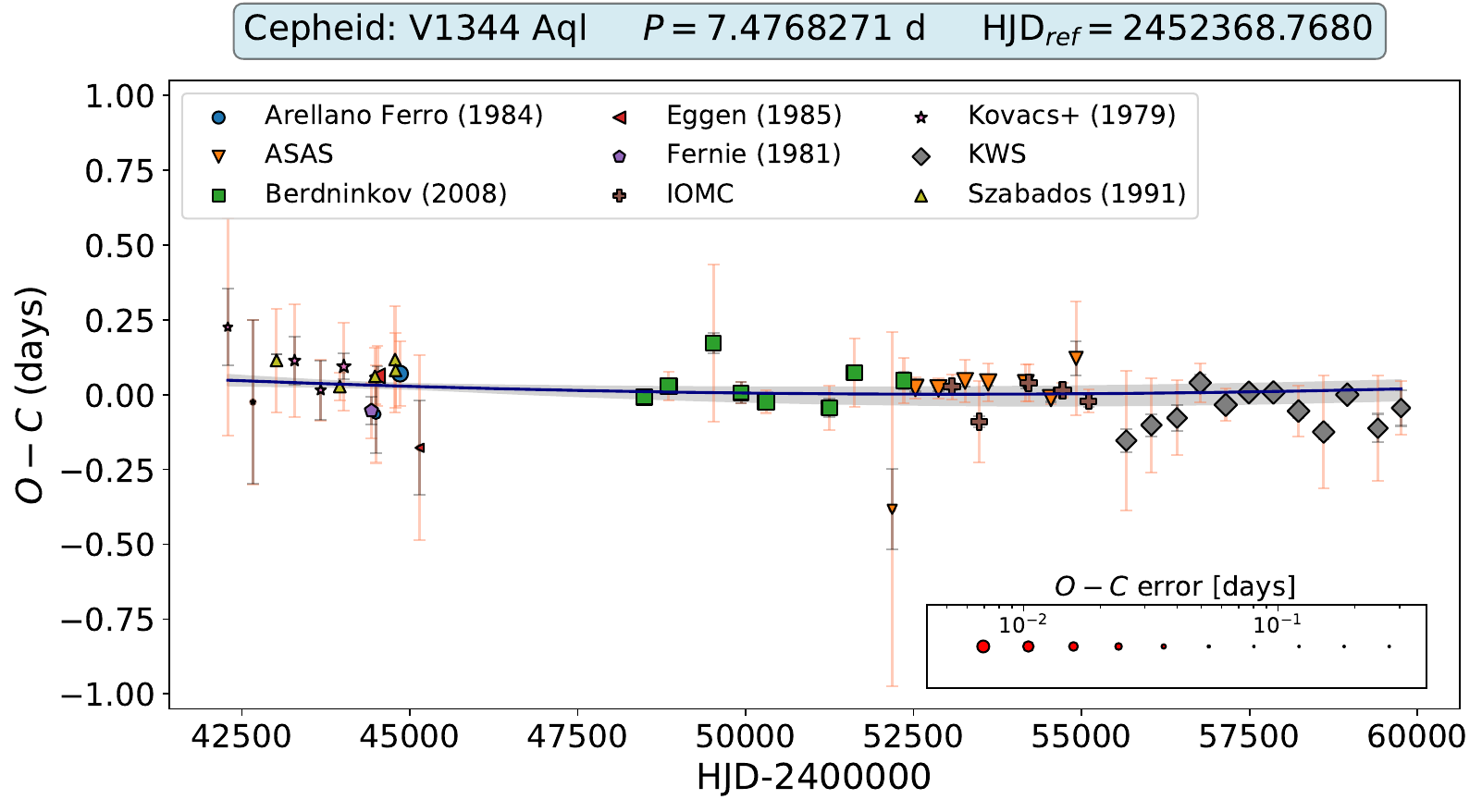}
    \caption{Updated $O-C$ diagram of V1344~Aql. The blue curve shows the best parabolic fit we obtained. The black error bars show the uncertainties as estimated by the \cite{Csornyei2022} pipeline, while the red ones show the inflated errors according to the \cite{Hogg2010} definition. In the inset, the plotted point size refers to the corresponding uncertainty.}
    \label{fig:o-c}
\end{figure*}

Figure~\ref{fig:o-c} shows the obtained $O-C$ diagram. The data shown in the plot are presented in Table~\ref{tab:OC}. The final data set utilised for deriving the $O-C$ values included observations from \cite{Arellano1984}, the All-Sky Automated Survey (ASAS, \citealt{ASAS}), \cite{Berdnikov2008}, \cite{Eggen1985}, \cite{Fernie1981}, the Integral Optical Monitoring Camera (IOMC, \citealt{IOMC}), \cite{Kovacs1979}, the KWS \citep{KWS} and \cite{Szabados1991}. To account for evolutionary changes and calculate the period values relevant to the times of the spectral observations, we fit a parabolic trend to the diagram. Given the extremely small scale of the period change and the relatively large scatter of points, we performed this fitting in a Bayesian framework accounting for the potentially underestimated errors and outliers following \cite{Hogg2010}. 

Based on the fit, following the methodology of \cite{Sterken2005}, the change in the pulsation period for V1344~Aql is 

\begin{equation}
    \dot{P} = 2.286\cdot 10^{-4} \pm 1.630\cdot 10^{-4}\,{\rm d/(100~yr)}.
\end{equation}

The estimated period change rate is $\sim$10$\%$ higher than the \cite{Csornyei2022} results, however, they are perfectly consistent within uncertainties. This estimate was used to infer the pulsation period at the chosen reference epoch in Eq.~\ref{eq:oc}. According to the earlier analysis by \cite{Csornyei2022}, the pulsation period of V1344~Aql at the above reference epoch was estimated as $P = 7.476713 \pm 0.000004$ days, compared to which the new value is 0.002 per cent longer. The offset between the previous and new period estimates is due to the more advanced treatment of uncertainties and the scatter among the data through the Bayesian fitting. Owing to the high fits uncertainties, as it was pointed out by \citet{Szabados14} as well, it is not possible to reliably separate the light-time effect caused by the orbital motion from the general scatter of the $O-C$ points. This issue might be remedied by the future publication of the complete Harvard photometric plates archive \citep{DASCH}, which will allow for the construction of a century-long $O-C$ diagram along with a more precise view of the period changes.

\section{New radial velocity data and orbit determination}
\label{sec:rvdata}

Additionally to the data set presented in the discovery paper on spectroscopic binarity of V1344~Aql by \citet{Szabados14}, new RV data were collected using the 0.5\,m RC telescope of ELTE Gothard Astrophysical Observatory, Szombathely and the 1\,m RCC telescope of Piszk\'estet\H{o} Mountain Station of the Konkoly Observatory of the Research Centre for Astronomy and Earth Sciences. 
The spectra were taken with two different fiber-fed echelle spectrographs both installed on optical telescopes: the eShel system of the French Shelyak Instruments \citep[$R \approx 11000$]{thizy} used on both telescopes before 2014 and the spectrograph made by Astronomical Consultants \& Equipment (ACE), having a resolution of $R \lessapprox 20000$ and mounted at the 1\,m RCC telescope since 2014. The exposure times were between 600\,s and 1200\,s depending on weather conditions, seeing and the used telescope. 
Th-Ar calibration spectra were taken before and after each observed target star in all cases.
The data reduction was done using the same method as described in \citet{Szabados14}, i.e. by applying standard tasks in \texttt{IRAF}\footnote{\texttt{IRAF} is distributed by the National Optical Astronomy Observatories, which are operated by the Association of Universities for Research in Astronomy, Inc., under cooperative agreement with the National Science Foundation.}, including bias, dark and flat-field corrections, aperture extraction, wavelength calibration, and continuum normalization. 
The consistency of the wavelength calibrations was checked using RV standard star observations, which proved the stability of the systems. 
The individual RV values were calculated using the cross-correlation technique with synthetic spectra from the \citet{munari2005} library with resolution $R$ = 11500, and effective temperature $T_{\mathrm{eff}}$ = 6000 K. 
We performed the calculations using the FXCOR task of \texttt{IRAF} in the wavelength range between 490 and 650 nm, excluding the wavelength range between 588 and 590 nm and 627 and 629 nm. 
Thus we do not include Balmer lines, NaD and telluric regions in the cross-correlation.
To calculate barycentric Julian dates and velocity corrections the \texttt{BARCOR} code of \citet{Hrudkova} was used for the mid-exposure times of the observations.

\begin{figure}[!ht]
\centering
\includegraphics[width=88mm]{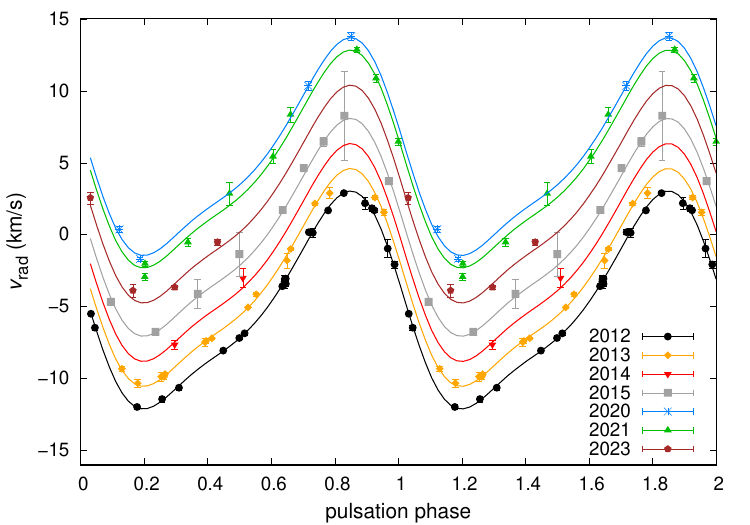}
\caption{Our RV data and the template curves for each year based on the fit to the data from 2012. The template curves are shifted vertically according to the $\gamma$-velocities given in Table~\ref{tab:vgamma}.}
\label{fig:rv_data}
\end{figure}

The RV curves were calculated using the period and epoch given in Section~\ref{sec:period}.
In order to achieve the best consistency, we refolded our previously published data \citep{Szabados14} to a phase curve using the new period.
The folded phase curves are shown in Figure~\ref{fig:rv_data}, while the new RV values along with their uncertainties (calculated by fitting a parabola around the maximum value of the cross-correlation function) are listed in Table~\ref{tab:rv_data}.
We used the RV observations from 2012 to construct a template curve for determination of the $\gamma$-velocity (velocity of the mass centre) of the Cepheid. 
We fitted a 3rd order Fourier-polynomial to the data, the $\gamma$-velocities were then calculated for each observing season (year) and are presented in Table~\ref{tab:vgamma}.
Since we used the period determined for 2020, no phase shift was necessary to fit the template curves and calculate the $\gamma$-velocities for each season. 
The template curves shown in Figure~\ref{fig:rv_data} represent the $\gamma$-velocity variation of the system, by shifting the curve vertically from 2012 with the appropriate $\gamma$-velocities given in Table~\ref{tab:vgamma}.

\begin{table}[!ht]
\caption{New RV data for V1344~Aql. }
\centering
\begin{tabular}{lrc}
\hline
BJD & $v_{\mathrm{rad}}$ (km s$^{-1}$) & err (km s$^{-1}$)\\
\hline
2456909.3875 &	$-$7.6	& 0.3 \\
2456918.4673 &	$-$3.0	& 0.7 \\
2457241.4049 &	4.6		& 0.2 \\
2457242.3617 &	8.3		& 3.1 \\
2457243.4088 &	3.7		& 0.2 \\
2457244.3442 &	$-$4.7	& 0.1 \\
2457245.3936 &	$-$6.8	& 0.2 \\
2457246.3873 &	$-$4.1	& 1.0 \\
2457247.3684 &	$-$1.4	& 1.5 \\
2457248.3865 &	1.7		& 0.1 \\
2457249.3502 &	6.5		& 0.3 \\
2459029.4833 &	13.8		& 0.3 \\
2459031.5060 &	0.4		& 0.2 \\
2459028.4748 &	10.4		& 0.3 \\
2459054.4271 &	$-$1.7    & 0.1 \\
2459379.4656 &  8.3       & 0.5 \\
2459383.5193 &	$-$2.0	& 0.2 \\
2459384.5282 &	$-$0.5	& 0.3 \\
2459385.5068 &	2.9		& 0.8 \\
2459386.5316 &	5.4		& 0.5	\\
2459388.5049 &	12.9		& 0.2 \\
2459389.4808 &	6.5		& 0.3 \\
2459411.3869 &	10.9		& 0.3 \\
2459413.4216 &	$-$2.9	& 0.2 \\
2460107.4857 &  2.6	    & 0.4 \\
2460108.4864 &  $-$3.9    & 0.5 \\
2460109.4707 &  $-$3.6    & 0.1 \\
2460110.4768 &  $-$0.5    & 0.2 \\
\hline
\end{tabular}
\label{tab:rv_data}
\end{table}

\begin{table}[!ht]
\centering
\caption{$\gamma$-velocities for V1344~Aql. $N$ marks the number of RV data points.}
\begin{tabular}{lccc}
\hline
mid-BJD & $v_{\gamma}$ (km s$^{-1}$) & $N$ & source \\
\hline
2444449 & $-$2.5 $\pm$ 0.5 & 24 & 1 \\
2444832 & +0.3 $\pm$ 0.5 & 8 & 2 \\
2456113 & $-$6.21 $\pm$ 0.02 & 22 & 3 \\
2456523 & $-$4.65 $\pm$ 0.03 & 16 & 3 \\
2456914 & $-$2.91 $\pm$ 0.19 & 2 & 3 \\
2457245 & $-$1.16 $\pm$ 0.07 & 9 & 3 \\
2459041 & +4.45 $\pm$ 0.11 & 4 & 3 \\
2459393 & +3.57 $\pm$ 0.09 & 9 & 3 \\
2460109 & +1.15 $\pm$ 0.13 & 4 & 3 \\
\hline
\end{tabular}
\tablebib{(1) \citet{balona1981}; (2) \citet{Arellano1984}; (3) present paper.}
\label{tab:vgamma}
\end{table}

\begin{figure}[!ht]
\centering
\includegraphics[width=88mm]{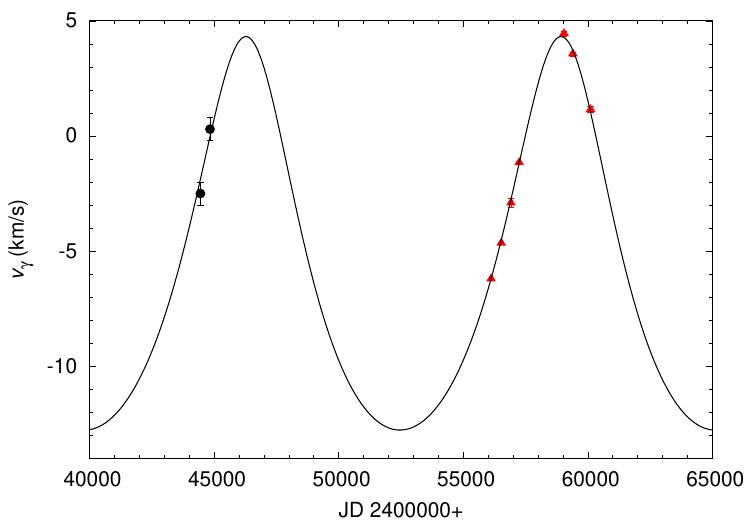}
\caption{Orbital solution with the parameters given in Table~\ref{tab:v1344aql_orbparam}. Our data are marked with red triangles, while black dots represent the earlier literature data (see Table \ref{tab:vgamma}).}
\label{fig:orbit}
\end{figure}

Our new RV data allowed us to determine the orbital solution for the system.
The fitting process was performed in two steps: first the grid search method \citep{Bevington} was applied using larger step sizes on a wider parameter space to find the location of the lowest $\chi^2$ values.
This method limited the location of the global minimum of the fit. 
In the second step the gradient search method \citep{Bevington} with $\chi^2$ minimisation was applied with varying step sizes to further restrict the orbital solution. 
The uncertainties of the fitted parameters were estimated using the bootstrap method \citep{bootstrap} by calculating one thousand orbits with the gradient search method. 
Finally, the standard deviation of the one thousand solutions is given as the uncertainty for each parameter. 
Our orbital solution is given in Table~\ref{tab:v1344aql_orbparam} and shown in Figure~\ref{fig:orbit}. 
We found that the orbital period of the system is approximately 34.6 years with an orbital eccentricity of 0.22.

\begin{table}[!ht]
\centering
\caption{New orbital parameters for V1344~Aql.}
\label{tab:v1344aql_orbparam}
\begin{tabular}{lc}
\hline
orbital element & value \\
\hline
 $a \sin i$ ($10^6$~km) & 1450 $\pm$ 16 \\
 $e$ & 0.22 $\pm$ 0.09 \\
 $\omega$ (rad) & 0.08 $\pm$ 0.08 \\
 $T_0$ (BJD) & 2459025 $\pm$ 43 \\
 $P_{\mathrm{orb}}$ (days) & 12649 $\pm$ 42 \\
 $v_0$ (km s$^{-1}$) & $-$6.1 $\pm$ 0.06 \\
 $K$ (km s$^{-1}$) & 8.55 $\pm$ 0.22 \\
 $f(m)$ (\msun) & 0.76 $\pm$ 0.02 \\
\hline
\end{tabular}
\tablefoot{The solution is drawn as a continuous line in Figure~\ref{fig:orbit}.}
\end{table}

\section{Discussion and conclusions}
\label{sec:summary}

Using our new RV data and updated pulsation period we were able to determine the orbital solution for the binary Cepheid V1344~Aql. 
We constructed an updated $O-C$ diagram based on the most recent observations to derive the recent pulsation period of the Cepheid. 

Since the temporal coverage of the observations does not allow us to separate the tiny light-time effect from the scatter of the $O-C$ values, we used only RV observations for the orbit determination.
A larger number of RV observations per observing season gives a lower uncertainty in the $\gamma$-velocities, however, a lower number of RV data covering the minimum and maximum phase are also useful for orbit determination. 
This can be seen in comparing our observations from 2020 and 2014: the four data points from 2020 include points close to the minimum and maximum velocity, thus the $\gamma$-velocity has a much lower uncertainty compared to that of 2014, where the two data points do not cover these phases.
The template curve fitting for 2014 still gives a $\gamma$-velocity that can be used for the orbit determination, although resulting in a larger uncertainty.

The maximum orbital velocity is close to the $\gamma$-velocity derived from our 2020 RV data, which has a huge impact on constraining the orbital solution.
We examined the possibility of a period approximately half of the one given in Table \ref{tab:v1344aql_orbparam}, that could not be excluded without our latest data point.
However, the $\chi^2$ value of this solution (including our latest point) was about 90 higher compared to the longer period orbit, without matching the maximum velocity phase, thus we excluded this solution as a possible fit.
The orbital period of the system is approximately 34.6 years with a relatively small eccentricity of 0.22.
This confirms the finding of \citet{Szabados14}, suggesting a decades-long orbit for the system.

RV and photometric observations in the forthcoming years will be crucial to disentangle and take into account the light-time effect of the system and to determine more accurate orbital elements.

\begin{acknowledgements}
The authors thank the referee for his/her constructive comments that helped to improve the quality of the paper.
This project has been supported by the Lend\"ulet grant LP2012-31 of the Hungarian Academy of Sciences. A. P\'al acknowledges support from the grant K-138962 of the Hungarian Research, Development and Innovation Office (NKFIH). LK acknowledges the support from the grants PD-134784, K-131508 and KKP-143986 of the Hungarian Research, Development and Innovation Office (NKFIH). LK is a Bolyai J\'anos Research Fellow. Financial support by the Hungarian National Research, Development and Innovation Office grant K129249 is acknowledged.
\end{acknowledgements}

%
%

\bibliographystyle{aa}
\bibliography{bibliography}

\begin{thebibliography}{36}
\expandafter\ifx\csname natexlab\endcsname\relax\def\natexlab#1{#1}\fi

\bibitem[{{Arellano Ferro}(1984)}]{Arellano1984}
{Arellano Ferro}, A. 1984, \mnras, 209, 481

\bibitem[{{Balona}(1981)}]{balona1981}
{Balona}, L.~A. 1981, The Observatory, 101, 205

\bibitem[{{Berdnikov}(2008)}]{Berdnikov2008}
{Berdnikov}, L.~N. 2008, VizieR Online Data Catalog, II/285

\bibitem[{{Bevington} \& {Robinson}(2003)}]{Bevington}
{Bevington}, P. \& {Robinson}, D., eds. 2003, Data Reduction and Error Analysis
  for the Physical Sciences (McGraw-Hill)

\bibitem[{{Breuval} {et~al.}(2020){Breuval}, {Kervella}, {Anderson}, {Riess},
  {Arenou}, {Trahin}, {M{\'e}rand}, {Gallenne}, {Gieren}, {Storm}, {Bono},
  {Pietrzy{\'n}ski}, {Nardetto}, {Javanmardi}, \& {Hocd{\'e}}}]{Breuval20}
{Breuval}, L., {Kervella}, P., {Anderson}, R.~I., {et~al.} 2020, \aap, 643,
  A115

\bibitem[{{Cs{\"o}rnyei} {et~al.}(2022){Cs{\"o}rnyei}, {Szabados},
  {Moln{\'a}r}, {Cseh}, {Egei}, {Kalup}, {Kecskem{\'e}thy},
  {K{\"o}nyves-T{\'o}th}, {S{\'a}rneczky}, \& {Szak{\'a}ts}}]{Csornyei2022}
{Cs{\"o}rnyei}, G., {Szabados}, L., {Moln{\'a}r}, L., {et~al.} 2022, \mnras,
  511, 2125

\bibitem[{{Derekas} {et~al.}(2012){Derekas}, {Szab{\'o}}, {Berdnikov},
  {Szab{\'o}}, {Smolec}, {Kiss}, {Szabados}, {Chadid}, {Evans}, {Kinemuchi},
  {Nemec}, {Seader}, {Smith}, \& {Tenenbaum}}]{Derekas2012}
{Derekas}, A., {Szab{\'o}}, G.~M., {Berdnikov}, L., {et~al.} 2012, \mnras, 425,
  1312

\bibitem[{Efron \& Tibshirani(1986)}]{bootstrap}
Efron, B. \& Tibshirani, R. 1986, Statist. Sci., 1, 54

\bibitem[{{Eggen}(1985)}]{Eggen1985}
{Eggen}, O.~J. 1985, \aj, 90, 1297

\bibitem[{{Evans} {et~al.}(2013){Evans}, {Bond}, {Schaefer}, {Mason},
  {Karovska}, \& {Tingle}}]{Evans13}
{Evans}, N.~E., {Bond}, H.~E., {Schaefer}, G.~H., {et~al.} 2013, \aj, 146, 93

\bibitem[{{Evans}(1992)}]{Evans92}
{Evans}, N.~R. 1992, \apj, 389, 657

\bibitem[{{Evans} {et~al.}(2015){Evans}, {Berdnikov}, {Lauer}, {Morgan},
  {Nichols}, {G{\"u}nther}, {Gorynya}, {Rastorguev}, \& {Moskalik}}]{Evans15}
{Evans}, N.~R., {Berdnikov}, L., {Lauer}, J., {et~al.} 2015, \aj, 150, 13

\bibitem[{{Fernie} \& {Garrison}(1981)}]{Fernie1981}
{Fernie}, J.~D. \& {Garrison}, R.~G. 1981, \pasp, 93, 330

\bibitem[{{Gaia Collaboration} {et~al.}(2017){Gaia Collaboration},
  {Clementini}, {Eyer}, {Ripepi}, {Marconi}, {Muraveva}, {Garofalo}, {Sarro},
  {Palmer}, {Luri}, {Molinaro}, {Rimoldini}, {Szabados}, {Musella}, {Anderson},
  {Prusti}, {de Bruijne}, {Brown}, {Vallenari}, {Babusiaux}, {Bailer-Jones},
  {Bastian}, {Biermann}, {Evans}, {Jansen}, {Jordi}, {Klioner}, {Lammers},
  {Lindegren}, {Mignard}, {Panem}, {Pourbaix}, {Randich}, {Sartoretti},
  {Siddiqui}, {Soubiran}, {Valette}, {van Leeuwen}, {Walton}, {Aerts},
  {Arenou}, {Cropper}, {Drimmel}, {H{\o}g}, {Katz}, {Lattanzi}, {O'Mullane},
  {Grebel}, {Holland}, {Huc}, {Passot}, {Perryman}, {Bramante}, {Cacciari},
  {Casta{\~n}eda}, {Chaoul}, {Cheek}, {De Angeli}, {Fabricius}, {Guerra},
  {Hern{\'a}ndez}, {Jean-Antoine-Piccolo}, {Masana}, {Messineo}, {Mowlavi},
  {Nienartowicz}, {Ord{\'o}{\~n}ez-Blanco}, {Panuzzo}, {Portell}, {Richards},
  {Riello}, {Seabroke}, {Tanga}, {Th{\'e}venin}, {Torra}, {Els},
  {Gracia-Abril}, {Comoretto}, {Garcia-Reinaldos}, {Lock}, {Mercier},
  {Altmann}, {Andrae}, {Astraatmadja}, {Bellas-Velidis}, {Benson}, {Berthier},
  {Blomme}, {Busso}, {Carry}, {Cellino}, {Cowell}, {Creevey}, {Cuypers},
  {Davidson}, {De Ridder}, {de Torres}, {Delchambre}, {Dell'Oro}, {Ducourant},
  {Fr{\'e}mat}, {Garc{\'\i}a-Torres}, {Gosset}, {Halbwachs}, {Hambly},
  {Harrison}, {Hauser}, {Hestroffer}, {Hodgkin}, {Huckle}, {Hutton},
  {Jasniewicz}, {Jordan}, {Kontizas}, {Korn}, {Lanzafame}, {Manteiga},
  {Moitinho}, {Muinonen}, {Osinde}, {Pancino}, {Pauwels}, {Petit},
  {Recio-Blanco}, {Robin}, {Siopis}, {Smith}, {Smith}, {Sozzetti}, {Thuillot},
  {van Reeven}, {Viala}, {Abbas}, {Abreu Aramburu}, {Accart}, {Aguado},
  {Allan}, {Allasia}, {Altavilla}, {{\'A}lvarez}, {Alves}, {Andrei}, {Anglada
  Varela}, {Antiche}, {Antoja}, {Ant{\'o}n}, {Arcay}, {Bach}, {Baker},
  {Balaguer-N{\'u}{\~n}ez}, {Barache}, {Barata}, {Barbier}, {Barblan}, {Barrado
  y Navascu{\'e}s}, {Barros}, {Barstow}, {Becciani}, {Bellazzini}, {Bello
  Garc{\'\i}a}, {Belokurov}, {Bendjoya}, {Berihuete}, {Bianchi},
  {Bienaym{\'e}}, {Billebaud}, {Blagorodnova}, {Blanco-Cuaresma}, {Boch},
  {Bombrun}, {Borrachero}, {Bouquillon}, {Bourda}, {Bragaglia}, {Breddels},
  {Brouillet}, {Br{\"u}semeister}, {Bucciarelli}, {Burgess}, {Burgon},
  {Burlacu}, {Busonero}, {Buzzi}, {Caffau}, {Cambras}, {Campbell},
  {Cancelliere}, {Cantat-Gaudin}, {Carlucci}, {Carrasco}, {Castellani},
  {Charlot}, {Charnas}, {Chiavassa}, {Clotet}, {Cocozza}, {Collins},
  {Costigan}, {Crifo}, {Cross}, {Crosta}, {Crowley}, {Dafonte}, {Damerdji},
  {Dapergolas}, {David}, {David}, {De Cat}, {de Felice}, {de Laverny}, {De
  Luise}, {De March}, {de Souza}, {Debosscher}, {del Pozo}, {Delbo}, {Delgado},
  {Delgado}, {Di Matteo}, {Diakite}, {Distefano}, {Dolding}, {Dos Anjos},
  {Drazinos}, {Dur{\'a}n}, {Dzigan}, {Edvardsson}, {Enke}, {Evans}, {Eynard
  Bontemps}, {Fabre}, {Fabrizio}, {Falc{\~a}o}, {Farr{\`a}s Casas}, {Federici},
  {Fedorets}, {Fern{\'a}ndez-Hern{\'a}ndez}, {Fernique}, {Fienga}, {Figueras},
  {Filippi}, {Findeisen}, {Fonti}, {Fouesneau}, {Fraile}, {Fraser}, {Fuchs},
  {Gai}, {Galleti}, {Galluccio}, {Garabato}, {Garc{\'\i}a-Sedano}, {Garralda},
  {Gavras}, {Gerssen}, {Geyer}, {Gilmore}, {Girona}, {Giuffrida}, {Gomes},
  {Gonz{\'a}lez-Marcos}, {Gonz{\'a}lez-N{\'u}{\~n}ez}, {Gonz{\'a}lez-Vidal},
  {Granvik}, {Guerrier}, {Guillout}, {Guiraud}, {G{\'u}rpide},
  {Guti{\'e}rrez-S{\'a}nchez}, {Guy}, {Haigron}, {Hatzidimitriou}, {Haywood},
  {Heiter}, {Helmi}, {Hobbs}, {Hofmann}, {Holl}, {Holland}, {Hunt}, {Hypki},
  {Icardi}, {Irwin}, {Jevardat de Fombelle}, {Jofr{\'e}}, {Jonker}, {Jorissen},
  {Julbe}, {Karampelas}, {Kochoska}, {Kohley}, {Kolenberg}, {Kontizas},
  {Koposov}, {Kordopatis}, {Koubsky}, {Krone-Martins}, {Kudryashova},
  {Bachchan}, {Lacoste-Seris}, {Lanza}, {Lavigne}, {Le Poncin-Lafitte},
  {Lebreton}, {Lebzelter}, {Leccia}, {Leclerc}, {Lecoeur-Taibi}, {Lemaitre},
  {Lenhardt}, {Leroux}, {Liao}, {Licata}, {Lindstr{\o}m}, {Lister}, {Livanou},
  {Lobel}, {L{\"o}ffler}, {L{\'o}pez}, {Lorenz}, {MacDonald}, {Magalh{\~a}es
  Fernandes}, {Managau}, {Mann}, {Mantelet}, {Marchal}, {Marchant}, {Marinoni},
  {Marrese}, {Marschalk{\'o}}, {Marshall}, {Mart{\'\i}n-Fleitas}, {Martino},
  {Mary}, {Matijevi{\v{c}}}, {McMillan}, {Messina}, {Michalik}, {Millar},
  {Miranda}, {Molina}, {Molinaro}, {Moln{\'a}r}, {Moniez}, {Montegriffo},
  {Mor}, {Mora}, {Morbidelli}, {Morel}, {Morgenthaler}, {Morris}, {Mulone},
  {Narbonne}, {Nelemans}, {Nicastro}, {Noval}, {Ord{\'e}novic},
  {Ordieres-Mer{\'e}}, {Osborne}, {Pagani}, {Pagano}, {Pailler}, {Palacin},
  {Palaversa}, {Parsons}, {Pecoraro}, {Pedrosa}, {Pentik{\"a}inen}, {Pichon},
  {Piersimoni}, {Pineau}, {Plachy}, {Plum}, {Poujoulet}, {Pr{\v{s}}a},
  {Pulone}, {Ragaini}, {Rago}, {Rambaux}, {Ramos-Lerate}, {Ranalli}, {Rauw},
  {Read}, {Regibo}, {Reyl{\'e}}, {Ribeiro}, {Riva}, {Rixon}, {Roelens},
  {Romero-G{\'o}mez}, {Rowell}, {Royer}, {Ruiz-Dern}, {Sadowski}, {Sagrist{\`a}
  Sell{\'e}s}, {Sahlmann}, {Salgado}, {Salguero}, {Sarasso}, {Savietto},
  {Schultheis}, {Sciacca}, {Segol}, {Segovia}, {Segransan}, {Shih},
  {Smareglia}, {Smart}, {Solano}, {Solitro}, {Sordo}, {Soria Nieto}, {Souchay},
  {Spagna}, {Spoto}, {Stampa}, {Steele}, {Steidelm{\"u}ller}, {Stephenson},
  {Stoev}, {Suess}, {S{\"u}veges}, {Surdej}, {Szegedi-Elek}, {Tapiador},
  {Taris}, {Tauran}, {Taylor}, {Teixeira}, {Terrett}, {Tingley}, {Trager},
  {Turon}, {Ulla}, {Utrilla}, {Valentini}, {van Elteren}, {Van Hemelryck}, {van
  Leeuwen}, {Varadi}, {Vecchiato}, {Veljanoski}, {Via}, {Vicente}, {Vogt},
  {Voss}, {Votruba}, {Voutsinas}, {Walmsley}, {Weiler}, {Weingrill}, {Wevers},
  {Wyrzykowski}, {Yoldas}, {{\v{Z}}erjal}, {Zucker}, {Zurbach}, {Zwitter},
  {Alecu}, {Allen}, {Allende Prieto}, {Amorim}, {Anglada-Escud{\'e}},
  {Arsenijevic}, {Azaz}, {Balm}, {Beck}, {Bernstein}, {Bigot}, {Bijaoui},
  {Blasco}, {Bonfigli}, {Bono}, {Boudreault}, {Bressan}, {Brown}, {Brunet},
  {Bunclark}, {Buonanno}, {Butkevich}, {Carret}, {Carrion}, {Chemin},
  {Ch{\'e}reau}, {Corcione}, {Darmigny}, {de Boer}, {de Teodoro}, {de Zeeuw},
  {Delle Luche}, {Domingues}, {Dubath}, {Fodor}, {Fr{\'e}zouls}, {Fries},
  {Fustes}, {Fyfe}, {Gallardo}, {Gallegos}, {Gardiol}, {Gebran}, {Gomboc},
  {G{\'o}mez}, {Grux}, {Gueguen}, {Heyrovsky}, {Hoar}, {Iannicola}, {Isasi
  Parache}, {Janotto}, {Joliet}, {Jonckheere}, {Keil}, {Kim}, {Klagyivik},
  {Klar}, {Knude}, {Kochukhov}, {Kolka}, {Kos}, {Kutka}, {Lainey}, {LeBouquin},
  {Liu}, {Loreggia}, {Makarov}, {Marseille}, {Martayan}, {Martinez-Rubi},
  {Massart}, {Meynadier}, {Mignot}, {Munari}, {Nguyen}, {Nordlander},
  {O'Flaherty}, {Ocvirk}, {Olias Sanz}, {Ortiz}, {Osorio}, {Oszkiewicz},
  {Ouzounis}, {Park}, {Pasquato}, {Peltzer}, {Peralta}, {P{\'e}turaud},
  {Pieniluoma}, {Pigozzi}, {Poels}, {Prat}, {Prod'homme}, {Raison}, {Rebordao},
  {Risquez}, {Rocca-Volmerange}, {Rosen}, {Ruiz-Fuertes}, {Russo}, {Serraller
  Vizcaino}, {Short}, {Siebert}, {Silva}, {Sinachopoulos}, {Slezak}, {Soffel},
  {Sosnowska}, {Strai{\v{z}}ys}, {ter Linden}, {Terrell}, {Theil}, {Tiede},
  {Troisi}, {Tsalmantza}, {Tur}, {Vaccari}, {Vachier}, {Valles}, {Van Hamme},
  {Veltz}, {Virtanen}, {Wallut}, {Wichmann}, {Wilkinson}, {Ziaeepour}, \&
  {Zschocke}}]{Gaia17}
{Gaia Collaboration}, {Clementini}, G., {Eyer}, L., {et~al.} 2017, \aap, 605,
  A79

\bibitem[{{Grindlay} {et~al.}(2012){Grindlay}, {Tang}, {Los}, \&
  {Servillat}}]{DASCH}
{Grindlay}, J., {Tang}, S., {Los}, E., \& {Servillat}, M. 2012, in IAU
  Symposium, Vol. 285, New Horizons in Time Domain Astronomy, ed. E.~{Griffin},
  R.~{Hanisch}, \& R.~{Seaman}, 29--34

\bibitem[{{Hogg} {et~al.}(2010){Hogg}, {Bovy}, \& {Lang}}]{Hogg2010}
{Hogg}, D.~W., {Bovy}, J., \& {Lang}, D. 2010, arXiv e-prints, arXiv:1008.4686

\bibitem[{{Hrudkov\'a}(2006)}]{Hrudkova}
{Hrudkov\'a}, M. 2006, in WDS'06 Proceedings of Contributed Papers: Part III -
  Physics, ed. J.~Safrankova \& J.~Pavlu (Matfyzpress, Prague), 18

\bibitem[{{Karczmarek} {et~al.}(2023){Karczmarek}, {Hajdu}, {Pietrzy{\'n}ski},
  {Gieren}, {Narloch}, {Smolec}, {Wiktorowicz}, \& {Belczynski}}]{Karczmarek23}
{Karczmarek}, P., {Hajdu}, G., {Pietrzy{\'n}ski}, G., {et~al.} 2023, \apj, 950,
  182

\bibitem[{{Karczmarek} {et~al.}(2022){Karczmarek}, {Smolec}, {Hajdu},
  {Pietrzy{\'n}ski}, {Gieren}, {Narloch}, {Wiktorowicz}, \&
  {Belczynski}}]{Karczmarek22}
{Karczmarek}, P., {Smolec}, R., {Hajdu}, G., {et~al.} 2022, \apj, 930, 65

\bibitem[{{Kervella} {et~al.}(2019){Kervella}, {Gallenne}, {Remage Evans},
  {Szabados}, {Arenou}, {M{\'e}rand}, {Proto}, {Karczmarek}, {Nardetto},
  {Gieren}, \& {Pietrzynski}}]{Kervella19}
{Kervella}, P., {Gallenne}, A., {Remage Evans}, N., {et~al.} 2019, \aap, 623,
  A116

\bibitem[{{Kovacs} \& {Szabados}(1979)}]{Kovacs1979}
{Kovacs}, G. \& {Szabados}, L. 1979, Information Bulletin on Variable Stars,
  1719, 1

\bibitem[{{Moe} \& {Di Stefano}(2017)}]{Moe17}
{Moe}, M. \& {Di Stefano}, R. 2017, \apjs, 230, 15

\bibitem[{{Morokuma} {et~al.}(2014){Morokuma}, {Tominaga}, {Tanaka}, {Mori},
  {Matsumoto}, {Kikuchi}, {Shibata}, {Sako}, {Aoki}, {Doi}, {Kobayashi},
  {Maehara}, {Matsunaga}, {Mito}, {Miyata}, {Nakada}, {Soyano}, {Tarusawa},
  {Miyazaki}, {Nakata}, {Okada}, {Sarugaku}, {Richmond}, {Akitaya}, {Aldering},
  {Arimatsu}, {Contreras}, {Horiuchi}, {Hsiao}, {Itoh}, {Iwata}, {Kawabata},
  {Kawai}, {Kitagawa}, {Kokubo}, {Kuroda}, {Mazzali}, {Misawa}, {Moritani},
  {Morrell}, {Okamoto}, {Pavlyuk}, {Phillips}, {Pian}, {Sahu}, {Saito}, {Sano},
  {Stritzinger}, {Tachibana}, {Taddia}, {Takaki}, {Tateuchi}, {Tomita},
  {Tsvetkov}, {Ui}, {Ukita}, {Urata}, {Walker}, \& {Yoshii}}]{KWS}
{Morokuma}, T., {Tominaga}, N., {Tanaka}, M., {et~al.} 2014, \pasj, 66, 114

\bibitem[{{Munari} {et~al.}(2005){Munari}, {Sordo}, {Castelli}, \&
  {Zwitter}}]{munari2005}
{Munari}, U., {Sordo}, R., {Castelli}, F., \& {Zwitter}, T. 2005, \aap, 442,
  1127

\bibitem[{{Neilson} {et~al.}(2015){Neilson}, {Schneider}, {Izzard}, {Evans}, \&
  {Langer}}]{Neilson15}
{Neilson}, H.~R., {Schneider}, F. R.~N., {Izzard}, R.~G., {Evans}, N.~R., \&
  {Langer}, N. 2015, \aap, 574, A2

\bibitem[{{Pilecki} {et~al.}(2022){Pilecki}, {Thompson}, {Espinoza-Arancibia},
  {Anderson}, {Gieren}, {Narloch}, {Minniti}, {Pietrzy{\'n}ski}, {Taormina},
  {Bono}, \& {Hajdu}}]{Pilecki22}
{Pilecki}, B., {Thompson}, I.~B., {Espinoza-Arancibia}, F., {et~al.} 2022,
  \apjl, 940, L48

\bibitem[{{Pojmanski}(2001)}]{ASAS}
{Pojmanski}, G. 2001, in Astronomical Society of the Pacific Conference Series,
  Vol. 246, IAU Colloq. 183: Small Telescope Astronomy on Global Scales, ed.
  B.~{Paczynski}, W.-P. {Chen}, \& C.~{Lemme}, 53

\bibitem[{{Sterken}(2005)}]{Sterken2005}
{Sterken}, C. 2005, in Astronomical Society of the Pacific Conference Series,
  Vol. 335, The Light-Time Effect in Astrophysics: Causes and cures of the O-C
  diagram, ed. C.~{Sterken}, 3

\bibitem[{{Sza\-bados} {et~al.}(2014){Sza\-bados}, {Cseh}, {Kov{\'a}cs},
  {Cs{\'a}k}, {D{\'o}zsa}, {Szab{\'o}}, {Simon}, {Borkovits}, {Kiss},
  {Jankovics}, \& {Mez{\H{o}}}}]{Szabados14}
{Sza\-bados}, L., {Cseh}, B., {Kov{\'a}cs}, J., {et~al.} 2014, \mnras, 442,
  3155

\bibitem[{{Sza\-bados} {et~al.}(2013){Sza\-bados}, {Derekas}, {Kiss},
  {Kov{\'a}cs}, {Anderson}, {Kiss}, {Szalai}, {Sz{\'e}kely}, \&
  {Christiansen}}]{Szabados13}
{Sza\-bados}, L., {Derekas}, A., {Kiss}, L.~L., {et~al.} 2013, \mnras, 430,
  2018

\bibitem[{{Szabados}(1991)}]{Szabados1991}
{Szabados}, L. 1991, Communications of the Konkoly Observatory Hungary, 96, 123

\bibitem[{{Szabados}(2003)}]{Szabados03}
{Szabados}, L. 2003, Information Bulletin on Variable Stars, 5394, 1

\bibitem[{{Szabados} \& {Klagyivik}(2012)}]{SzabadosKlagyivik12}
{Szabados}, L. \& {Klagyivik}, P. 2012, \apss, 341, 99

\bibitem[{{Szabados} \& {Neh{\'e}z}(2012)}]{Szabados12_LMC}
{Szabados}, L. \& {Neh{\'e}z}, D. 2012, \mnras, 426, 3148

\bibitem[{{Thizy} \& {Cochard}(2011)}]{thizy}
{Thizy}, O. \& {Cochard}, F. 2011, in Proceedings of the IAU, Vol. 272, Active
  OB stars: structure, evolution, mass loss, and critical limits, ed.
  {C.~Neiner, G.~Wade, G.~Meynet, \& G.~Peters}, 282

\bibitem[{{Winkler} {et~al.}(2003){Winkler}, {Courvoisier}, {Di Cocco},
  {et~al.}}]{IOMC}
{Winkler}, C., {Courvoisier}, T.~J.~L., {Di Cocco}, G., {et~al.} 2003,
  Astronomy \& Astrophysics, 411, L1

\end{thebibliography}

\begin{appendix}
\section{$O-C$ data}
Table.~\ref{tab:OC} shows the median $V$-band $O-C$ values obtained for V1344~Aql. The underlying data is identical to that analysed in \cite{Csornyei2022}, except for the most recent $KWS$ observations. 

\begin{table}[!ht]
\centering
\caption{Obtained $O-C$ values along with the source data references. }
\begin{tabular}{l r r r c }
\hline
HJD & Epoch & $O-C_{\textrm{med}}$ & $O-C_{\textrm{err}}$ & Reference \\
\hline
2442290.2052  & $-$2229.0 & 0.226 & 0.068 & 1 \\
2442663.7960  & $-$2179.0 & $-$0.025 & 0.271 & 1 \\
2443016.8960  & $-$2132.0 & 1.664 & 0.030 & 2\\
2443284.5110  & $-$2096.0 & 0.114 & 0.081 & 1 \\
2443673.2077  & $-$2044.0 & 0.015 & 0.084 & 1 \\
2443958.8890  & $-$2006.0 & 1.577 & 0.030 & 2\\
2444017.2203  & $-$1998.0 & 0.094 & 0.056 & 1 \\
2444482.3010  & $-$1936.0 & 1.611 & 0.030 & 2\\
2444495.5779  & $-$1934.0 & $-$0.065 & 0.126 & 3 \\
2444518.1365  & $-$1931.0 & 0.063 & 0.030 & 4 \\
2444781.4300  & $-$1896.0 & 1.667 & 0.030 & 2\\
2444788.8710  & $-$1895.0 & 1.631 & 0.030 & 2\\
2444854.6011  & $-$1886.0 & 0.070 & 0.022 & 3 \\
2445145.9497  & $-$1847.0 & $-$0.177 & 0.175 & 4 \\
2444428.2985  & $-$1943.0 & $-$0.053 & 0.041 & 5 \\
2448495.7383  & $-$1399.0 & $-$0.007 & 0.024 & 6 \\
2448854.6619  & $-$1351.0 & 0.029 & 0.021 & 6 \\
2449520.2433  & $-$1262.0 & 0.173 & 0.038 & 6 \\
2449931.3028  & $-$1207.0 & 0.007 & 0.031 & 6 \\
2450305.1143  & $-$1157.0 & $-$0.023 & 0.015 & 6 \\
2451247.1728  & $-$1031.0 & $-$0.045 & 0.022 & 6 \\
2451621.1321  & $-$981.0 & 0.073 & 0.021 & 6 \\
2452181.4378  & $-$906.0 & $-$0.383 & 0.147 & 7 \\
2452353.8345  & $-$883.0 & 0.047 & 0.027 & 6 \\
2452525.7775  & $-$860.0 & 0.023 & 0.013 & 7 \\
2452869.7095  & $-$814.0 & 0.021 & 0.012 & 7 \\
2453071.5898  & $-$787.0 & 0.027 & 0.011 & 8 \\
2453266.0061  & $-$761.0 & 0.046 & 0.018 & 7 \\
2453475.2214  & $-$733.0 & $-$0.090 & 0.017 & 8 \\
2453609.9357  & $-$715.0 & 0.041 & 0.013 & 7 \\
2454170.6954  & $-$640.0 & 0.039 & 0.018 & 7 \\
2454215.5582  & $-$634.0 & 0.041 & 0.008 & 8 \\
2454544.4862  & $-$590.0 & $-$0.012 & 0.013 & 7 \\
2454716.4797  & $-$567.0 & 0.015 & 0.007 & 8 \\
2454918.4603  & $-$540.0 & 0.121 & 0.046 & 7 \\
2455105.2383  & $-$515.0 & $-$0.021 & 0.020 & 8 \\
2455665.8685  & $-$440.0 & $-$0.153 & 0.036 & 9 \\
2456039.7613  & $-$390.0 & $-$0.102 & 0.034 & 9 \\
2456421.1034  & $-$339.0 & $-$0.078 & 0.041 & 9 \\
2456765.1552  & $-$293.0 & 0.040 & 0.029 & 9 \\
2457146.3993  & $-$242.0 & $-$0.034 & 0.016 & 9 \\
2457490.3746  & $-$196.0 & 0.007 & 0.018 & 9 \\
2457856.7392  & $-$147.0 & 0.007 & 0.016 & 9 \\
2458230.5185  & $-$97.0 & $-$0.055 & 0.023 & 9 \\
2458604.2897  & $-$47.0 & $-$0.125 & 0.029 & 9 \\
2458955.8253  & 0.0 & 0.000 & 0.016 & 9 \\
2459411.7998  & 61.0 & $-$0.112 & 0.055 & 9 \\
2459755.8012  & 107.0 & $-$0.045 & 0.074 & 9 \\
\hline
\end{tabular}
\tablefoot{The $O-C$ values and their errors are measured in days.}
\tablebib{
 (1) \cite{Kovacs1979}; (2) \cite{Szabados1991}; (3) \cite{Arellano1984}; (4) \cite{Eggen1985}; (5) \cite{Fernie1981}; (6) \cite{Berdnikov2008}; (7) ASAS, \cite{ASAS}; (8) IOMC, \cite{IOMC}; (9) KWS, \cite{KWS}.}
\label{tab:OC}
\end{table}

\end{appendix}

\end{document}